\documentclass[12pt,preprint]{aastex}

\newcommand{\kpc} {$\, {\rm kpc}$}
\newcommand{\kms} {$\, {\rm km/s}$}

\newcommand{\vect}[1]{\ensuremath{\mbox{\boldmath $#1$}}}

\begin{document}

\def\newpage{\vfill\eject}
\def\vs{\vskip 0.2truein}
\def\pp{\parshape 2 0.0truecm 16.25truecm 2truecm 14.25truecm}
\def\fun#1#2{\lower3.6pt\vbox{\baselineskip0pt\lineskip.9pt
  \ialign{$\mathsurround=0pt#1\hfil##\hfil$\crcr#2\crcr\sim\crcr}}}
\def\core{{\rm core}}
\def\min{{\rm min}}
\def\max{{\rm max}}
\def\kpc{{\rm kpc}}
\def\esc{{\rm esc}}
\def\crit{{\rm crit}}
\def\pc{{\rm pc}}
\def\kms{{\rm km}\,{\rm s}^{-1}}
\def\cbh{{\rm cbh}}
\def\bh{{\rm bh}}
\def\df{{\rm df}}
\def\bulge{{\rm bulge}}

\title{Millisecond Pulsars as Probes of Mass Segregation in the Galactic Center}

\author{Julio Chanam\'e\, and Andrew Gould} 
\affil{{}Department of Astronomy, The Ohio State University,
Columbus, OH 43210, USA}

\begin{abstract}
We propose a simple test for the existence of a cluster of black hole
remnants around Sgr A* that is based on a small sample of any type of
Galactic Center objects, provided they are substantially less massive
than the black holes and constitute part of an old ($\gtrsim$ 1 Gyr)
population.  The test relies on the fact that, under the presence of
such a cluster of heavy remnants and because of energy equipartition,
lower mass objects would be expelled from the central regions and
settle into a distribution very different than the cusp expected to be
induced by the supermassive black hole alone.  We show that with a
sample of just 50 objects and using only their angular positions on
the sky relative to Sgr A* it is possible to clearly differentiate
between a distribution consistent with the presence of the cluster of
black holes and a power-law cusp distribution.  We argue that
millisecond pulsars might currently be the best candidate to perform
this test, because of the large uncertainties involved in the age
determination of less exotic objects.  In addition, by measuring their
first and second period derivatives, millisecond pulsars offer the
rare opportunity of determining the complete phase space information
of the objects.  We show that this extra information improves the
detection of mass segregation by about 30\%.
\end{abstract}

\setcounter{footnote}{0}
\renewcommand{\thefootnote}{\arabic{footnote}}

\section{Introduction}

Measurements of the proper motions, radial velocities, and even the
accelerations of the closest resolvable stars at the Galactic Center
(GC) have demonstrated the existence of a large concentration of mass
inside their orbits, leading to the almost inevitable conclusion that
a black hole with a mass of $M_{\ast}\approx 3 \times 10^{6}$
M$_{\odot}$ lies at the center of the Galaxy, coinciding with the
position of the GC radio source Sgr A* (Eckart \& Genzel 1997; Ghez et
al.\ 1998; Genzel et al.\ 2000; Ghez et al.\ 2000).  The same
improvements in near infrared (NIR) detection limits and in angular
resolution that made this kind of work possible also show that the
stars around the GC are part of a complex mixture of old and young
populations, unlike the familiar, old Galactic bulge.  Massive and
young associations such as the Arches and Quintuplet clusters (Figer
et al.\ 1999), HeI emission-line stars (Najarro et al. 1997) as well
as Wolf-Rayet stars (Blum, Sellgren \& Depoy 1995; Figer, McLean \&
Morris 1995), all represent clear evidence of episodes of star
fomation near the GC during the last 10$^{7}$ years (Krabbe et al.\
1995).

In dynamical terms, a supermassive black hole is expected to produce a
power law distribution, $\rho(r) \sim r^{-\alpha}$, of stars inside a
cusp radius $r_{c}$ on a timescale comparable to the relaxation
timescale at $r_{c}$.  Considering the case of a massive black hole in
the center of globular clusters, Bahcall \& Wolf (1976) predicted the
formation of a $r^{-7/4}$ cusp for a population of stars of equal
masses.  For a more realistic multi-mass stellar distribution it was
found that each mass species forms a different cusp with power-law
index in the range $3/2\leq\alpha\leq 7/4$ (Bahcall \& Wolf 1977;
Murphy et al.\ 1991), with the heavier stars more concentrated towards
the center than the lighter ones.  Evidence for this mass segregation
in globular clusters exists in various forms: the slope of the stellar
mass function (as derived from the luminosity function at different
radii) continuously decreasing towards the cluster core, as found with
the recent {\it Hubble Space Telescope (HST)} data on 47 Tuc (Howell
et al.\ 2001; see also Sosin 1997); a more centrally concentrated
radial distribution of objects such as pulsars (Phinney 1992) and blue
stragglers (Cot\'e, Richer \& Fahlman 1991; Lanzeral et al.\
1992; Layden et al.\ 1999; Marconi et al. 2001) as compared to
subgiants and main sequence stars; and radial color gradients (Howell,
Guhathakurta \& Tan 2000).  Measurement of the surface brightness
profiles of the innermost parts of elliptical galaxies have shown that
almost all of them, both ``power-law'' and ``cuspy core'' galaxies in
the Nuker-type nomenclature, exhibit power law profiles all the way
down to the $\sim 0.\arcsec1$ resolution of {\it HST} (Faber et al.\
1997; van der Marel 1999).  Finally, in the context of the Galaxy and
based on analysis of the star counts from recent NIR observations of
the GC, Alexander (1999) concluded that the distribution of stars
around Sgr A* reveals the presence of a cusp with a power-law index
consistent with the Bahcall \& Wolf (1977) solution, arguing that a
flat core is completely ruled out and that the apparent overdensity of
faint stars in the inner $\sim$ 0.05 pc (the Sgr A* cluster) is the
tip of an underlying stellar cusp that smoothly rises throughout the
inner 10$\arcsec$.  However, most of the stars so far detected in the
NIR imaging of the GC are part of the youngest stellar population, and
hence they are dynamically unrelaxed, i.e., they have not had enough
time to achieve energy equipartition with the fainter, older stellar
population that dominates the mass in the inner Galaxy.  A detailed
theoretical investigation of the distribution of stars of different
masses around Sgr A* constitutes an interesting problem on its own and
will be addressed in a future paper.  Here we only want to note that
the Bahcall \& Wolf (1977) solution applies under assumptions of
steady state as well as specific boundary conditions, both of which do
not necessarily correspond to the conditions at the GC.

Recently, Miralda-Escud\'e \& Gould (2000) predicted the existence of
a cluster of black hole remnants around the GC.  Integrating over the
lifetime of the Galaxy and considering the rate of capture by Sgr A*,
they concluded that about 20,000 of these black holes have migrated by
dynamical friction to the GC and should still be there, forming a
compact cluster about 0.7 pc in radius ($\sim 18\arcsec$ for a
Galactocentric distance of 8 kpc).  Chanam\'e, Gould, \&
Miralda-Escud\'e (2001) considered the possibility of detecting the
cluster black holes by monitoring the pairs of images of background
bulge stars that are lensed by Sgr A*, looking for microlensing events
induced by the surrounding black holes.  They found that there should
be about 2 such pairs for $K<21$ mag, and about 8 down to $K$=23.
Then, if any of these pairs happens to be part of a high magnification
event, the rate of microlensing events due to the cluster black holes,
and hence the probability of detecting them, is reasonably high.

However, the simplest observable consequence of such a cluster of
black holes is that, as a result of relaxation, any sufficiently old
population of lower mass objects should have been expelled from the
region occupied by the black hole cluster.  This redistribution occurs
on a timescale of about 1 Gyr (the relaxation timescale at a radius of
0.7 pc), so this effect should only be apparent in populations that
are older than this (and substantially less massive than the black
holes).  Hence, the most direct way of testing for the presence of the
cluster of black holes is to measure the distribution of the oldest
possible population of objects and see whether they show this ``core''
or ``hole'' in their underlying distribution. 

We show that if any such population is identified, then with only
about 50 objects it is possible to differentiate between a
distribution driven by the presence of a cluster of black holes and a
regular power law distribution, by using only information on their
angular positions on the sky.

Current age determinations of stars at the GC are still concentrated
on the brightest objects, which are usually the most massive and
youngest.  Using $K$-band spectra, Blum, Sellgren \& DePoy (1996)
studied 19 bright GC stars (most of them supergiants and asymptotic
giant branch stars), deriving relatively accurate ages for the
youngest ones.  These stars are not expected to show any effect from
the cluster of black holes, both because their parent gas clouds do
not fall into the lower-mass category and also because, since they are
young, they have not been around long enough to be affected by the
process of relaxation.  Unfortunately, the uncertainties quoted by
Blum et al. (1996) for the oldest stars in their sample are so large
that they could place only lower limits on their ages ($\sim$ 440
Myr).  Until these uncertainties are greatly improved there will be no
hope of using ordinary stars for testing the mass segregation induced
by the cluster of black holes, because any contamination of the sample
with young objects would undermine one's ability to differentiate
between the two distributions.  The oldest objects possible, as well
as the most numerous, should be old, low-mass main sequence stars.
Reaching these stars at the distance of the GC requires a sensitivity
of at least $K\sim 21$ (the main sequence turnoff being at $K\sim
20$), which estimate takes into account an average of 3 magnitudes of
extinction in the $K$-band (see the empirical luminosity function of
the GC derived in Chanam\'e et al.\ 2001).  Moreover, at $K\sim 21$
one expects $\sim 400 \,{\rm arcsec}^{-2}$ stars of similar magnitude
and brighter, so that milliarcsec resolution would be required.  Such
deep, high-resolution observations are beyond current capabilities,
but may be achievable with improvements in adaptive optics, or with
NGST.

A very interesting alternative are millisecond (ms) pulsars.  Being
neutron stars, these are lighter than the remnant black holes and,
furthermore, constitute an old population of objects as estimated by
their spindown timescale $P/(2\dot{P})$, typically several Gyr for ms
pulsars, $\,\log(P_{\rm sec}) \lesssim -2$ (Taylor, Manchester, \&
Lyne 1993).  Since ms pulsars would be selected on an observable
directly related to their ages, the contamination by young objects
would be extremely low.  Moreover, pulsars offer the possibility of
adding more information than just the usual sky positions and proper
motions.  Measurements of the first and second period derivatives of a
pulsar orbiting the GC yield, up to a two-fold ambiguity, its position
and velocity along the line of sight, hence completing the entire
6-dimensional phase-space information.  Certainly, both $\dot{P}$ and
$\ddot{P}$ have contributions that are intrinsic to the pulsar.
However, for ms pulsars, $(\dot{P}/P)_{\rm int}\lesssim 10^{-17}\,
{\rm s}^{-1}$, is more than two orders of magnitude smaller than the
variation induced by the acceleration towards Sgr A* over the region
of interest, $(\dot{P}/P)_{GC} = (GM_{\ast}c^{-1})/(0.7 {\rm pc})^{2}
\approx 3\times 10^{-15}\, {\rm s}^{-1}$ (note that, 10 times farther away 
than this, i.e., 3 arcmin in projection from the GC, the intrinsic
$\dot{P}/P$ already becomes important).  Similarly, $(\ddot{P}/P)_{\rm
int}\sim 10^{-31}\, {\rm s}^{-2}$ (Phinney 1993), again, much smaller
than the contribution from the Sgr A* potential, $(\ddot{P}/P)_{GC} =
(\dot{P}/P)_{GC}/\tau_{orb} \approx 2\times 10^{-26}\, {\rm s}^{-2}$,
where $\tau_{orb}$ is the orbital timescale of the pulsar around the
GC.  Hence, the intrinsic period derivatives of the pulsars do not
represent a limitation for our purposes.  Nevertheless, because of the
crowded GC environment passing stars can produce effects that mimic
$\ddot{P}$, as we discuss in \S\,2.2.  We find that the extra
information obtained by measuring the position and velocity of the
pulsars along the line of sight does indeed improve one's ability to
differentiate between different underlying pulsar distributions as
compared with using only the sky positions.  However, the improvement
is modest, only $\sim$ 30\%.

A problem with using pulsars as probes of mass segregation arises
because the high density of free electrons in the central regions of
the Galaxy strongly scatters radio waves, broadening the pulsar pulses
by extraordinary amounts and hence making difficult their detection by
periodicity searches (see Cordes \& Lazio 1997).  As a result, today
there are no known pulsars closer than 1$^{\circ}$ to the GC.
Nevertheless, the degree of pulse broadening is strongly frequency
dependent ($\Delta t$ $\sim \nu^{-4}$), and most pulsar searches have
been made at frequencies like 0.4 GHz, where this effect is known to
be large.  Using a model for the distribution of free electrons in the
Galaxy (Taylor \& Cordes 1993), Johnston et al.\ (1995) carried a
search for pulsars near the GC at 1.5 GHz, a frequency specifically
chosen so as to minimize the broadening of the pulses ($\Delta t$(0.4
GHz)/$\Delta t$(1.5 GHz) = (1.5/0.4)$^{4}$ $\approx$ 200).  None were
found, suggesting that there is substantial scattering in the GC that
is not taken into account by the Taylor \& Cordes (1993) electron
density model.  It is clear then, that future searches for pulsars
near the GC must be done at even higher frequencies, such as 5 or 15
GHz, where the pulse broadening is smaller than at 1.5 GHz by factors
of 10$^{2}$ and 10$^{4}$, respectively.  Improvements in sensitivity
are also needed for such a search, both in order to have more
potential targets (the observed local luminosity function of pulsars
scales as $N(L) \sim L^{-1}$) and because the pulsar spectral energy
distribution decreases with frequency ($F_{\nu} \sim \nu^{-1.4}$).
The planned Square Kilometer Array (SKA), with about two orders of
magnitude increased sensitivity over existing facilities, is the
indicated choice for these searches.  Another possibility is to
conduct imaging rather than periodicity surveys.  These may be far
more successful since angular broadening is not as severe as pulse
broadening (see Cordes \& Lazio 1997).

In \S\,2 we present the input density models used to simulate data,
describe our maximum likelihood approach, and show how ms pulsars
improve the results as compared to less exotic objects.  In
\S\,3 we present the results and discuss them, summarizing our
conclusions.

\section{Input Density Profile and Likelihood Analysis}

\subsection{The General Case}

Lower mass objects expelled from the inner GC region will show in
their distribution an overdensity at some radius $r_{0}$ of the order
of the radius of the black hole cluster, producing something that
would resemble a ``core'' or ``hole'' around the GC.  We model this
with the family of functions

\begin{equation}
\nu(r) = C\,\biggl(\,\frac{1}{r^{2}}+\frac{1}{r_{0}^{2}}\,\biggr)^{-\alpha/2}\,\biggl(r^{2}+r_{0}^{2}\biggr)^{-\beta/2},
\end{equation}

\noindent and study combinations of ($r_{0},\alpha,\beta$).  The
larger the value of $\alpha$ the more evident the core/hole around the
GC, and the larger the value of $\beta$ the narrower the distribution
around $r_{0}$.  The normalization constant is chosen such that the
volume integral of $\nu(r)$ is equal to the size of the sample.  Note
that fixing $\alpha$ equal to zero and using a very small $r_{0}$,
yields a power-law density profile, $\nu(r) \sim r^{-\beta}$.  Figure
1 illustrates the differences between a distribution for the tracer
population that is consistent with a black hole cluster around the GC,
and a power law distribution.  Figure 1 shows the model with the
biggest hole considered in this work.

We first describe the simplest case, in which the only information
used is the pulsar positions on the plane of the sky.  Later more
information will be included.  Given a set of data
$\lbrace\vect{\theta}_{k};\,k=1...N\rbrace$ on $N$ objects (where
$\vect{\theta}_{k}=(\theta^{x}_{k},\theta^{y}_{k})$ represents the
angular position of the $k$-th object) that follow some unknown
underlying parent distribution, denoted as
($r_{0}^{*},\alpha^{*},\beta^{*}$), we want:

\begin{itemize} 

\item First, to compare the likelihoods that all those coordinates are 
coming from: (a) a distribution consistent with a cluster of black
holes being at the GC, and (b) a power-law distribution, as would be
expected in the absence of a black hole cluster.

\item Second, to know how accurately the underlying parameters
($r_{0}^{*},\alpha^{*},\beta^{*}$) can be recovered from the data.

\end{itemize}

\noindent To achieve this, we choose the input model
($r_{0}^{*},\alpha^{*},\beta^{*}$) consistent with the presence of a
cluster of black holes (i.e., with a core/hole in the distribution
around the GC) and generate random positions for $N$ objects.  This
data set is then given to a routine that, using the downhill SIMPLEX
method for minimization in multidimensions (Press et al.\ 1992), finds
the parameters of the distribution ($r_{0},\alpha,\beta$) that
maximize the likelihood function, as given by

\begin{equation}
{\rm ln}\, {\cal L} = \sum^{N}_{k=1}\, {\rm ln}\,{\cal P}(\,\lbrace\vect{\theta}_{k}
\rbrace\,\vert\, r_{0},\alpha,\beta\,)\, = \sum^{N}_{k=1}\, {\rm ln}\,\int^{\infty}_{-\infty}dz\,\,\nu(r_{k})dx_{k}dy_{k}\,\,,
\end{equation}

\noindent where ${\cal P}(\,\lbrace \vect{\theta}_{k}\rbrace\,\vert\, r_{0},
\alpha,\beta\,)$ denotes the probability of the ocurrence of the 
{\it k}-th data point given the model distribution
($r_{0},\alpha,\beta$) being tested,
$r_{k}=\sqrt{x_{k}^{2}+y_{k}^{2}+z^{2}}$, and the integral over $z$
reflects our (in general) ignorance of the position of any given
object along the line of sight.  Equation (2) makes immediately
obvious that, in order to transform from the observable quantities
$\lbrace\vect{\theta}_{k}\rbrace$ to actual lengths, one needs to
assume a distance to the GC, $D_{GC}$.  Throughout this paper we adopt
$D_{GC} = 8$ kpc.  We discuss the effects of the uncertainty in this
parameter in \S\,3.  The experiment is repeated several times (each
time using a new set of data) for the two distributions that we wish
to compare.  When testing for a power-law distribution, we fix the
value of $\alpha$ to zero and of $r_{0}$ to a very small value, 0.1.
Computing the statistics of the outcome of all these experiments we
finally obtain:

\begin{itemize}

\item the difference between the mean values of ln${\cal L}$ for the two 
competing distributions, which tells whether or not it is possible to
differentiate between them and with how much confidence.

\item the mean and variances of each of the fitted parameters 
($r_{0},\alpha,\beta$), which tells how well the input model
($r_{0}^{*},\alpha^{*},\beta^{*}$) can be recovered from the data.

\end{itemize}

Equation (2) illustrates the simplest case, in which the available
information reduces to just the two-dimensional positions of the
objects on the sky.  It is still possible to include two-dimensional
velocity information by measurement of the proper motions,
$\vect{\mu}_{k}=(\mu^{x}_{k},\mu^{y}_{k})$, and that would be, for a
general class of objects, all the information that one could possibly
add.

\subsection{Millisecond Pulsars}

However, if the objects being used for this test are ms pulsars, then
it is possible to add even more information to the likelihood, the
position and velocity of the pulsars along the line of sight ($z$ and
$v_{z}$).  This is accomplished as follows.  First, the line of sight
acceleration, $a_{z}$, can be derived by measuring the first time
derivative of the pulsar's period, $\dot{P}$, as has been succesfully
done for several pulsars in globular clusters (Phinney 1993; Robinson
et al.\ 1995; Freire et al.\ 2001).  This acceleration can in turn be
related to the actual pulsar position with respect to Sgr A* by

\begin{equation}
\frac{GM}{b^{2}}\,\sin^{2}\eta\,\cos\eta = -a_{z} = -\frac{\dot{P}}{P}\,c\,,
\end{equation}

\noindent where $M$ is the mass interior to the pulsar radius
(esentially the mass of Sgr A*), $b$ is the impact parameter with
respect to Sgr A*, and $\eta$ is the angle defined by the observer,
the pulsar, and Sgr A*.  It is necessary to stress here that this can
be safely done only in the case of ms pulsars, for which the
contribution to $\dot{P}$ due to intrinsic pulsar spindown is
negligible with respect to the contribution due to acceleration
towards Sgr A* over the region of interest.  The function
$\sin^{2}\eta\,\cos\eta$ is double valued in the interval [0,$\pi$],
so there are two possible values for the position along the line of
sight, $z = b\cot\eta$.  Both degenerate solutions must enter in the
calculation, since they represent a fundamental uncertainty that can
not be broken with the information considered here.  The measurement
of $\eta$ together with the position on the sky gives complete spatial
information, and the distance from Sgr A* is simply $r = b\csc\eta$.
Second, two components of the velocity, {\bf v$_{\perp}$}, are given
by measurement of the pulsar's proper motion.  The third component of
the velocity can be determined by measuring the pulsar jerk,
$\dot{a}_{z}$, from measurement of $\ddot{P}/P$.  Specifically,

\begin{equation}
v_{z} = \frac{b}{GM\,\sin\eta\,(3\cos^{2}\eta -1)}\,\biggl(\,\frac{\dot{a}_{z}b^{2}}{\sin^{2}\eta}+3a_{z}{\rm\bf b \cdot v_{\perp}}\biggr)\,,
\end{equation}

\noindent where the expression has been written in terms of
observables, except for the two-fold discretely degenerate parameter
$\eta$.  When using millisecond pulsars, then, the probability that
goes into equation (2) must be the sum of the individual probabilities
for each of the two degenerate pulsar positions, i.e., ${\cal
P}(\,\lbrace\vect{\theta}_{k},\vect{\mu}_{k},\dot{P}_{k},\ddot{P}_{k}
\rbrace\,\vert\, r_{0},\alpha,\beta\,) = {\cal P}_{k1}+{\cal P}_{k2}$,
where

\begin{equation}
{\cal P}_{ki} = \nu(r_{ki})\,\frac{1}{(2\pi\sigma_{ki}^{2})^{3/2}}\,\exp\biggl[-(v_{x}^{2}+v_{y}^{2}+v_{zi}^{2})/2\sigma_{i}^{2}\biggr]_{k}\,d^{6}V_{k}^{i}\,\,;\,\,i=1,2\,.
\end{equation}

\noindent Here $\sigma_{ki} = \sigma(r_{ki})$ is the one-dimensional
velocity dispersion at $r_{ki}$, computed assuming an isotropic
velocity distribution for the model $(r_{0},\alpha,\beta)$ (Binney \&
Tremaine 1987), and
$d^{6}V_{k}^{i}=dx_{k}\,dy_{k}\,dz_{k}^{i}\,dv_{xk}\,dv_{yk}\,dv_{zk}^{i}$
is the six-dimensional phase-space volume element centered on
$\lbrace\vect{x}_{i},\vect{v}_{i}\rbrace_{k}$.  Due to the
transformation from observables to actual positions and velocities,
this volume element is not the same for the two degenerate positions.
Hence, while the part $dx_{k}\,dy_{k}\,dv_{xk}\,dv_{yk}$ can be simply
factored-out from the expression for the probability (because it is
the same for both $i$=1,2), one must replace
\,$dz_{k}^{i}\,dv_{zk}^{i}\longrightarrow
J_{k}^{i}\,d\dot{P}_{k}\,d\ddot{P}_{k}$, where $J_{k}^{i}$ is the
Jacobian of the transformation
$(z,v_{z})_{k}\longrightarrow(\dot{P},\ddot{P})_{k}$.  Only now can
one factor out the observable part of the volume element, keeping the
factors $J_{k}^{i}$, which need to be computed each time for ${\cal
P}_{k1}$ and ${\cal P}_{k2}$.

The time derivatives of the pulsar's period, $\dot{P}$ and $\ddot{P}$,
are currently measured to great precision: for a 10 ms pulsar and over
a time baseline of 3 years, Phinney (1993) quotes a measurement
accuracy of $\dot{P}/P
\approx 3\times 10^{-20} {\rm s}^{-1}$, and $\ddot{P}/P \approx
10^{-27} {\rm s}^{-2}$.  Recall that, as discussed in \S\,1, the
pulsar's intrinsic period derivatives are not a source of uncertainty
for our purposes.  However, a passing star can introduce
uncertainties, though these are only important to the determination of
the pulsar's jerk.  The effect of a passing star of mass $m_{i}$ on
$\ddot{P}$ can be estimated computing the product of the induced
acceleration, $G\,m_{i}/d^{2}$, and $(\sqrt{3}\,\sigma_{stars}/d)$,
where $\sigma_{stars}$ is the one-dimensional velocity dispersion of
the stars at the pulsar's position and $d$ is the star-pulsar
distance, and asking what this distance would have to be in order for
the passing star to produce an effect as large as that due to Sgr A*,
$(\ddot{P}/P)_{GC}$.  We obtain, $d\backsimeq
(4/11)^{1/6}\,(m_{i}/M_{\ast})^{1/3}r$, where $r$ is the distance to
Sgr A* and the numerical factor in front comes from assuming a stellar
density profile proportional to $r^{-7/4}$.  Finally, the probability
of such a close encounter with a star of this particular type (mass
$m_{i}$) goes as $n_{i}\,d^{3}$, where $n_{i}$ is the number density
of stars of mass $m_{i}$.  Summing over all the types of stars, this
probability then scales as $\rho\, r^{3}$, where $\rho(r)$ is the
total stellar mass density, as given by equation (3) in
Miralda-Escud\'e \& Gould (2000).  Specifically, we obtain
$(4\pi/3)(12/11)^{1/2}\rho(r)\,r^{3}/M_{\ast}
\approx 0.13$, i.e., there is a modest chance that a close
encounter affects the pulsar's jerk by a large factor.  A similar
argument shows that the probability that a passing star will
significantly affect $\dot{P}$ is only $\sim 10^{-4}$ (see also
Phinney 1993).

\section{Results and Discussion}

We run simulations consisting of 400 experiments, where an experiment
includes both the generation of the data and fitting them to either
one of the competing distributions, as described in \S\,2.1.  In each
experiment we generate data for $N = 50$ objects.  Table 1 shows the
results of simulations for four different input models, labeled by the
corresponding combination of
$(r_{0}^{\ast},\alpha^{\ast},\beta^{\ast})$.  The models presented in
Table 1 are chosen to have varying sizes of the inner core/hole in the
distribution of stars, with the first one being the most evident (see
Fig. 1), peaking beyond the radius predicted for the cluster of black
holes ($\sim$ 0.7 pc, Miralda-Escud\'e \& Gould 2000).  Then, by using
smaller values for $r^{\ast}_{0}$ and $\alpha^{\ast}$, this hole is
made progressively smaller in the next two models, until in the third
model the peak is at about half the black hole cluster radius.  The
third model has a tail at large radii that falls slower than in the
first two models (i.e., smaller $\beta^{\ast}$), making the
distribution less concentrated around the peak, and hence even more
difficult to differentiate from a simple power law.  The fourth model
has $\alpha$=0.0, so that a core, rather than a hole, is set at the
origin.  Each entry in the table, which we denote $\Delta {\rm
ln}{\cal L}$, is the difference between the mean values of ln$\cal L$
when fitting for a distribution consistent with the presence of a
cluster of black holes and when fitting for a power law distribution,
{\it in that order}.  For each input model we compute $\Delta {\rm
ln}{\cal L}$ in four different cases (columns 2 to 5 in Table 1), each
one providing an increasing amount of information to the likelihood
calculation.  Note that columns 3 and 4 do not represent realistic
data sets (since they imply knowledge of the 3 dimensional position
and velocity of the star without any ambiguity), but they are helpful
in understanding both what kind of information is the most valuable
(i.e., that information that contributes the most to ln$\cal L$), and
to what degree the ambiguity in the $z$ position degrades the
likelihood.

\clearpage
\begin{deluxetable}{ccccccrrrrrrrrrr}
\footnotesize
\tablecaption{$\Delta {\rm ln}{\cal L}$ \label{tbl-1}}
\tablewidth{0pt}
\tablehead{
\colhead{input model} & \colhead{ $(x,y)$} & \colhead{$(x,y,z_{1})$} & \colhead{ ($\vect{x}_{1},\vect{v}_{1}$)} & \colhead{millisecond} \\ \colhead{$(r_{0}^{\ast},\alpha^{\ast},\beta^{\ast})$} & \colhead{} & \colhead{} & \colhead{} & \colhead{pulsars}
}
\startdata
(1.0, 2.0, 4.0)  &22.3  &27.1  &29.8  &28.1  \nl
(0.5, 1.0, 4.0)  &16.2  &20.0  &22.2  &20.9  \nl
(0.3, 0.5, 3.3)  &4.7   &5.8   &6.4   &6.0   \nl
(1.0, 0.0, 4.0)  &18.6  &22.5  &24.8  &23.0  \nl
\enddata



\end{deluxetable}
\clearpage

The first column of $\Delta {\rm ln}{\cal L}$ values (second column in
Table 1), labeled $(x,y)$, represents the simplest case (see eq. 2),
in which the only information available for the test is the pulsar
positions on the plane of the sky.  As expected, the larger the hole
at the center of the distribution, the larger the value of $\Delta
{\rm ln}{\cal L}$, i.e., the easier it is to distinguish between the
two competing models.  The conclusion from this first set of
experiments is that with the angular positions of only 50 GC objects
one can easily tell if their distribution is consistent or not with a
cluster of black holes at the GC, {\it provided that all the objects
in the sample are old enough {\rm ($\gtrsim$ 1 Gyr)} to have achieved
relaxation}.

Millisecond pulsars have the advantage that they do not present any
age related problem, and furthermore, as discussed in \S\,2.2, they
offer the unique opportunity to obtain the complete phase space
coordinates on the individual objects, so that all the space and
velocity information can be included in the likelihood calculation.
In the fifth (last) column of Table 1 we present the results of
experiments that simulate the use of ms pulsars, i.e., they assume
knowledge of all six phase space coordinates and take into account the
two-fold ambiguity described in \S\,2.2.  One can see that $\Delta
{\rm ln}{\cal L}$ grows in all four input models by about 30\% of the
value obtained when using only the positions on the sky.  We conclude
then that this extra information coming from the use of ms pulsars
does not improve the confidence in the results by a large factor: it
has the same effect as adding $\sim$ 15 objects to the original sample
and performing the test using just the angular positions on the sky.
Hence, the real advantage of the use of ms pulsars is the certainty
that one is using a sufficiently old population, so that young,
unrelaxed objects do not contaminate the sample.

With the aim of gaining a better understanding on the nature of the
information that is enhancing and/or degrading the value of $\Delta
{\rm ln}{\cal L}$, we perform two extra sets of experiments under the
(unrealistic) supposition that no ambiguity in the determination of
the three-dimensional position and velocity of the pulsars is present
at all.  First, we suppose that the three dimensional position,
$(x,y,z_{1})$, can be known without ambiguity, and second, we suppose
that the three dimensional velocity, $(v_{x},v_{y},v_{z1})$, can also
be known with no ambiguity.  The results of these extra experiments
are given in the third and fourth columns of Table 1, respectively.
Comparison between the fourth and fifth columns immediately shows that
the two-fold ambiguity in the actual determination of $z$ and $v_{z}$
hardly degrades $\Delta {\rm ln}{\cal L}$ at all relative to complete
phase information.  Instead, by comparing the second and third columns
one realizes that by far the largest improvement comes from knowledge
of the 3-d position in space with respect to Sgr A*.  Finally, adding
the velocity information (compare the third and fourth columns in
Table 1) makes only a marginal improvement.  Recall here, from the
discussion in \S\,2.2, that the line-of-sight velocity of the pulsars
are the most uncertain measurements of those considered here, so,
happily for our purposes, they are at the same time the least useful.
That is, the spatial position of the objects with respect to Sgr A* is
the most important ingredient in this test, and, fortunately, it
happens to be the easiest to determine.

Finally, what are the effects of an uncertainty in the value of
$D_{GC}$?  The distance to the GC enters our calculation in two
places: through the positions and velocities that are simulated, and
through $M_{\ast}$, the adopted total mass of the supermassive black
hole.  The latter scales as $\propto D_{GC}^{3}$ if it is determined
from measurements of the proper motions of the stars close to Sgr A*.
Hence, our likelihood analysis should not be affected when using only
angular positions and proper motions.  The coordinates along the line
of sight ($z$ and $v_{z}$), however, have a non trivial dependence on
$D_{GC}$, so it is not easy to predict the effect of varying $D_{GC}$
in the case where we use all phase space information.  In order to
quantify these we compute $\Delta {\rm ln}{\cal L}$ for different
values of $D_{GC}$.  We indeed find that when using only angular
positions, the value of $\Delta {\rm ln}{\cal L}$ remains exactly the
same regardless of the value of $D_{GC}$.  This is not the case when
using all the phase space information, for which a 10\% increase in
the adopted Galactocentric distance has the effect of changing the
value of $\Delta {\rm ln}{\cal L}$ by a small amount (about 2\% of
tabulated values), hence introducing no effective change in one's
ability to distinguish the two different distributions.  When
performing the test with only the angular positions as inputs (as in
the second column in Table 1), the recovered (fitted) {\it angular
size} of the hole in the distribution, $r_{0}/D_{GC}$, remains
constant, as expected when using just angular measurements.  However,
when including all the phase space information (as in the last column
in Table 1) the hole's angular size increases approximately linearly
with $D_{GC}$.  The fitted angles $\alpha$ and $\beta$, as well as the
uncertainties in the three parameters, do not change when varying
$D_{GC}$.  We conclude that the uncertainty in the Galactocentric
distance does not affect the reliability of the test in determining
the underlying distribution of the tracers.

In summary:

\begin{itemize}

\item the simple analysis of the angular distribution of any old 
population of objects that have achieved relaxation around the GC
constitutes a powerful probe of mass segregation in the GC, in
particular as a test for the existence of a cluster of stellar mass
black holes around Sgr A*.

\item a sample of only 50 objects is enough to obtain reliable results, 
and the only measurements needed are their positions relative to Sgr
A*.  Velocity information does not contribute appreciably to the
results.

\item provided it is older than $\sim$ 1 Gyr, any population of 
objects with masses substantially lower than that of the black holes
is in principle equally well fitted to be used in the test.  Being
intrinsically a very old population, ms pulsars might currently be the
most promising candidate, because the determination of the ages of
normal stars at the GC is too uncertain.  However, improvements in
search techniques, as well as in sensitivity at radio wavelengths, are
needed to find pulsars in the difficult GC environment.  Future radio
facilities such as the planned SKA will probably be required to carry
out the high frequency pulsar searches ($\nu \gtrsim$ 5 GHz) and/or
imaging surveys needed to find pulsars near the GC.

\item the use of ms pulsars provides the opportunity of determining 
the complete phase space information of the objects, improving the
results by about 30\%.  The improvement comes primarily from the
addition of information on the position $z$ of the pulsars along the
line of sight with respect to Sgr A*, i.e., from the measurement of
the pulsar's first period derivative (\S\,2.2).  Measurement of the
second period derivative leads to knowledge of the velocity along the
line of sight, but there is little to be gained from this.

\end{itemize}

%

\begin{acknowledgements}

Work by AG was supported in part by grant AST 97-27520 from the NSF.

\end{acknowledgements}






\clearpage

\begin{figure}
\vspace*{-1cm}
\plotone{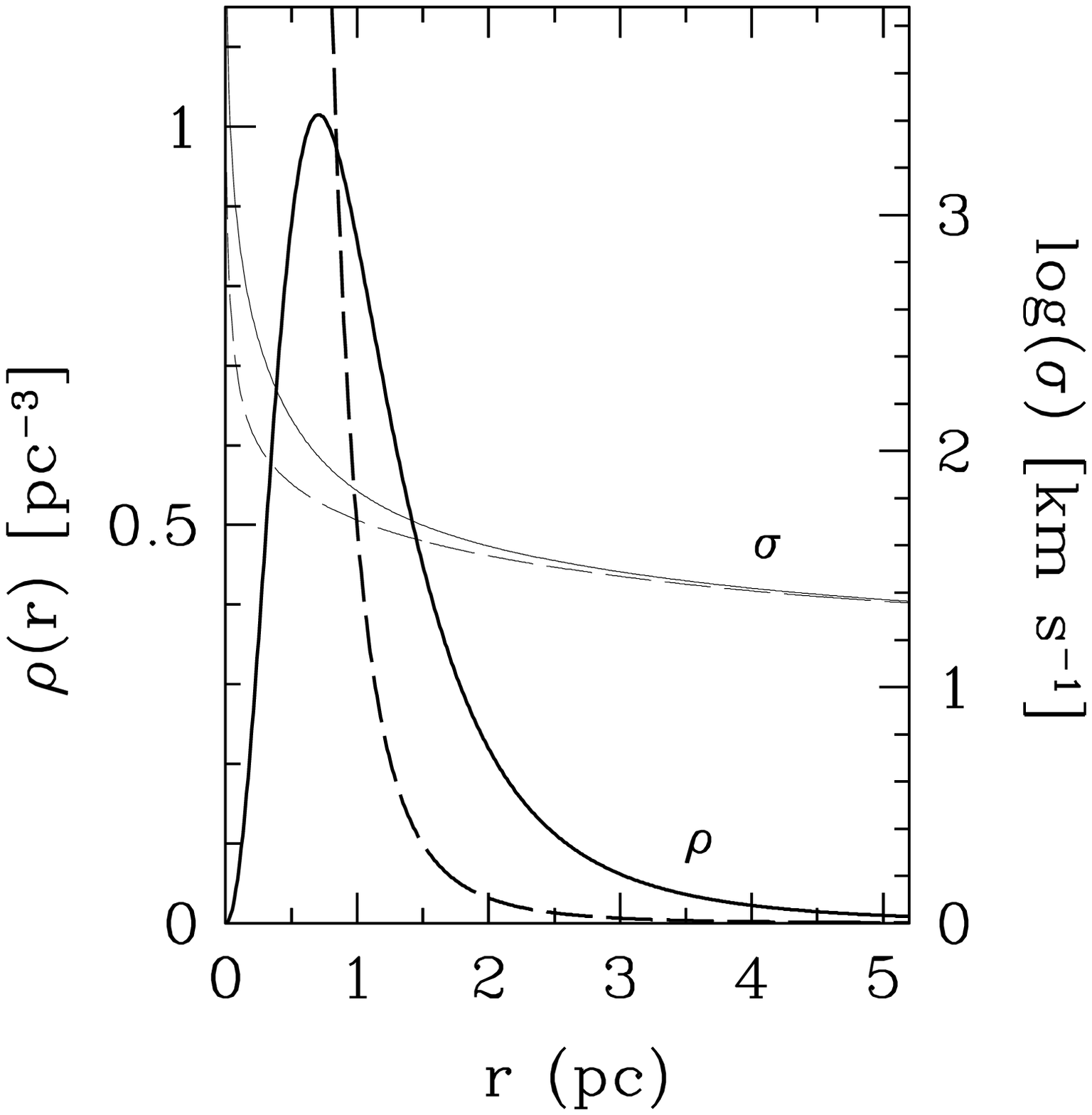}
\vspace*{-3cm}
\caption[junk]{\label{fig:one}
Illustration of the two competing distributions.  Solid lines
correspond to a distribution consistent with a cluster of black holes
around the GC, with parameters ($r_{0},\alpha,\beta$)=(1.0, 2.0, 4.0),
and dashed lines correspond to a power law distribution with
parameters ($r_{0},\alpha,\beta$)=(0.1, 0.0, 4.0).  }
\end{figure}

\end{document}